\documentclass[aip, apl, amsmath, amssymb, reprint]{revtex4-1}

\usepackage{graphicx}
\usepackage{dcolumn}
\usepackage{bm}

\begin{document}%

\title{Detection limits in whispering gallery biosensors with plasmonic enhancement}%

\author{Jon D. Swaim}%
\affiliation{Department of Physics, University of Queensland, St Lucia, QLD 4072 Australia}%
\author{Joachim Knittel}%
\affiliation{Department of Physics, University of Queensland, St Lucia, QLD 4072 Australia}%
\author{Warwick P. Bowen}%
 \email{wbowen@physics.uq.edu.au}%
\affiliation{Department of Physics, University of Queensland, St Lucia, QLD 4072 Australia}%
\affiliation{Centre for Engineered Quantum Systems, University of Queensland, St Lucia, QLD 4072, Australia}%

\date{14 December 2011}%

\begin{abstract}%
We perform numerical modeling of a gold nanorod bound to the surface of a microtoroid-based biosensor.  Localized surface plasmon resonances in the nanorod give rise to strong enhancements in the electric field when excited near resonance, increasing the frequency shift for a single bovine serum albumin molecule by a factor of 870, with even larger enhancements predicted for smaller proteins.  On resonance, the frequency shift is predicted to be on the order of MHz, more than an order of magnitude larger than measurement noise arising from time-averaged frequency and thermal fluctuations.
\end{abstract}%

\maketitle%

Whispering gallery mode (WGM) resonators such as silica (SiO$_2$) microspheres~\cite{Cai} and microtoroids~\cite{Vahala} have unprecedented sensitivity as biological sensors~\cite{Heavywater, Arnold, Knittel, Armani} due to their small optical mode volume and ultra-high quality factor ($Q>10^8$) in water.  The interaction between the resonator's evanescent field and its environment shifts the resonance frequency of the optical mode, such that when a molecule binds to the resonator surface it induces a shift given by~\cite{Arnold}
\begin{equation}
\frac{\delta \omega_m}{\omega} \simeq -\frac{\alpha_m |E_0(\vec{r}_m)|^2}{2 V |E_{0,\rm{max}}|^2}  
\label{eq:freq_shift}
\end{equation}
where $\alpha_m$ and $\vec{r}_m$ are the polarizability and position of the molecule, $E_0(\vec{r})$ is the WGM electric field and $V$ is the optical mode volume of the resonator.  For typical WGM resonators and proteins, Eq.~(\ref{eq:freq_shift}) predicts single molecule optical frequency shifts in the Hz to kHz range.  As an example, a bovine serum albumin (BSA; $\alpha_m=54800$ $\AA^3$)~\cite{Vollmer2} protein bound to a microtoroidal resonator with $V=760$ $\mu$m$^3$ produces a frequency shift of $\delta \omega_m=$ 5 kHz.  Presently, however, measurement noise arising primarily from thermal fluctuations within the resonator and fluctuations in laser frequency often limits the minimum detectable frequency shift to above 5 kHz~\cite{Armani, Lu}, precluding the observation of single molecule binding events.

Recently, it was shown experimentally that a localized surface plasmon resonance (LSPR) in a large metallic nanoshell can enhance single nanoparticle optical frequency shifts in a microsphere resonator by a factor of four~\cite{shopova}.  The enhancement was predicted to be as much as 200 for BSA.  A similar effect was shown in Ref.~\cite{Vollmer-NP}.  In this letter we predict that a gold (Au) nanorod with a length-to-diameter aspect ratio of $R=4$ can enhance the frequency shift of a single BSA molecule by as much as 870, with even larger enhancements possible for proteins smaller than BSA.  We compare the enhanced frequency shift for BSA with that of a bare microtoroid resonator, and show that the enhancement can allow single molecule detection below the limits set by thermal and laser frequency fluctuations.

\begin{figure}[hb!]
	\begin{center}
	\includegraphics*[scale=0.5]{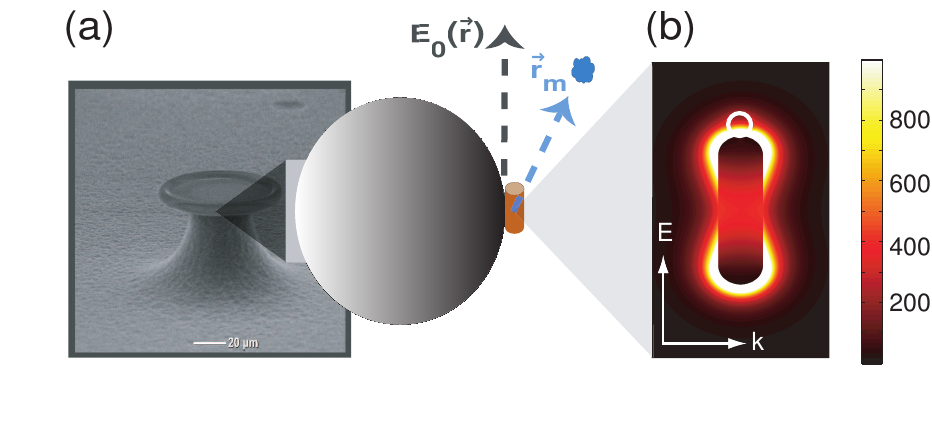}
	\caption{(a) An SEM image of a 70 $\mu$m microtoroid, and a side view of an equatorialy bound Au nanorod interacting with a target molecule (shown in blue).  (b)  Electric field intensity $|E_{nr}(\vec{r})|^2/|E_0(\vec{r})|^2$ around a 10 nm $\times$ 40 nm nanorod.}
	\label{fig:toroid}
	\end{center}
\end{figure}

We present theoretical modeling of the LSPR generated in Au nanorods using a Boundary Element Method (BEM) as described in Refs.~\cite{Abajo1, Abajo2, pt-nanorods}.  The model consists of a single Au nanorod bound to the equator of a microtoroid so that it experiences the maximum possible resonator evanescent field $E_0(\vec{r})$ (Fig.~\ref{fig:toroid} (a)), and orientated such that its longitudinal axis is aligned with the polarization of the WGM.  The nanorods' dimensions (diameter of 10 nm, $2 \leq R \leq 4$) are small in comparison with the incident wavelength $\lambda$, so the evanescent field may be modeled as a plane wave of unit amplitude, neglecting the small evanescent decay across the nanorod.  The background medium is water (with relative permittivity $\epsilon_b=1.77$), and the experimental dielectric function for Au was taken from Johnson and Christy~\cite{Johnson}.  First let us consider a bare nanorod without any target molecules present.  We will then consider the case when a single BSA molecule (white circle in Fig.~\ref{fig:toroid} (b)) binds to the tip of the nanorod and shifts the frequency of the microtoroid WGM.
\begin{figure}[ht!]
	\begin{center}
	\includegraphics*[scale=0.5]{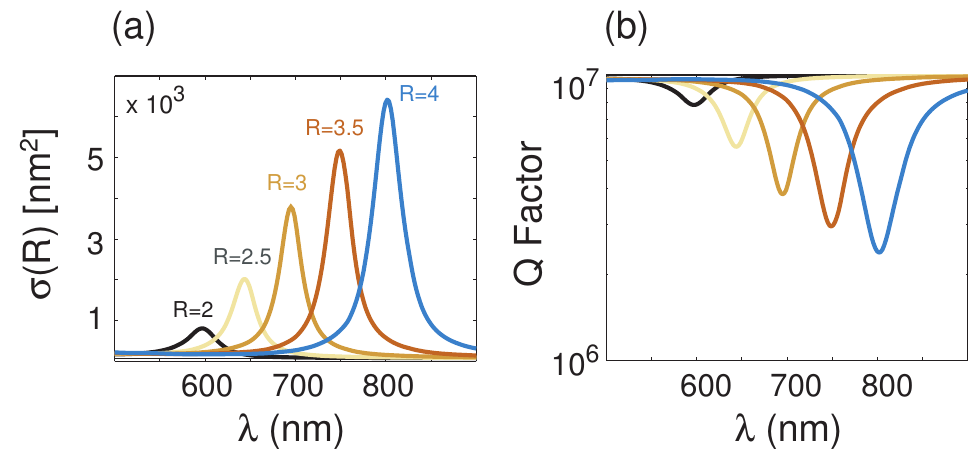}
	\caption{(a) Absorption cross-sections of nanorods with diameter = 10nm and varying aspect ratio $R$.  (b) WGM Q factor in the presence of nanorod absorption. Q$_0$=10$^7$, $V=760$ $\mu$m$^3$ and $f^2(\vec{r}_s)=0.3$.}
	\label{fig:nanorod_data}
	\end{center}
\end{figure}   

In Fig.~\ref{fig:toroid} (b) we show the electric field intensity around a nanorod with $R=4$ that is excited close to resonance ($\lambda =$ 803 nm).  The strong fields at each tip result from the curvature of the nanorod's hemispherical caps, with enhancements relative to the incident intensity in excess of $10^3$.  The enhancement has two important effects on the WGM: (1) the resonantly enhanced polarizability of the nanorod scatters light back in the microtoroid, resulting in doublet WGMs split in frequency, and (2) absorption from the nanorod will reduce the Q factor of the resonator.  Fig.~\ref{fig:nanorod_data} (a) shows the absorption cross-sections $\sigma(R)$ for nanorods with increasing $R$ (in order of increasing $\lambda$) calculated using the BEM.  To account for the enhanced absorption near resonance, we consider an additional cavity loss rate $\Gamma$ which reduces the optical Q factor of the resonator from $Q_0$ to $Q = (Q_0^{-1}+\Gamma / \omega)^{-1}$.  Absorption dominates over scattering in the near-field~\cite{Kreibig}, so the loss rate can be determined by $I_{\mathrm{inc}} \sigma = \hbar \omega \Gamma$, where $I_{\mathrm{inc}}=\hbar \omega c f^2(\vec{r}_s) / V$ is the intensity incident on the nanorod~\cite{Mazzei}, $c$ is the speed of light and $f(\vec{r}_s)$ is the spatial variation of $E_0(\vec{r})$ in the optical mode~\cite{Yang}.  The Q factor is then
\begin{equation}
Q= \left(Q_0^{-1} + \frac{\sigma(R) \lambda}{2 \pi V } f^2(\vec{r}_s)\right)^{-1}
\label{eq:retard}
\end{equation}
The reduced WGM Q factors are shown in Fig.~\ref{fig:nanorod_data} (b) for Q$_0=10^7$, $V=760$ $\mu$m$^3$ and $f^2(\vec{r}_s)=0.3$ at the equator.

\begin{figure}[hb!]
	\begin{center}
	\includegraphics*[scale=0.5]{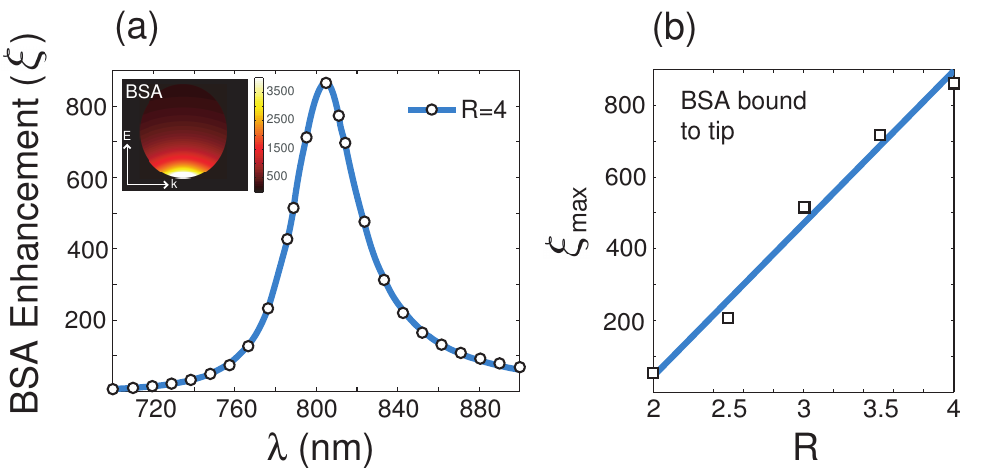}
	\caption{(a) Frequency shift enhancement for a BSA molecule due to a nanorod with $R=4$.  Inset shows a non-uniform intensity  distribution $|E_m(\vec{r})|^2/|E_0(\vec{r})|^2$ within the BSA molecule.  (b)  Maximum enhancement as a function of aspect ratio $R$.}
	\label{fig:enhancement}
	\end{center}
\end{figure}

Now we can consider the enhancement in $\delta \omega$ due to a BSA molecule binding to the microtoroid-bound nanorod.  We model the BSA molecule as a dielectric sphere with a radius of 3 nm and $\epsilon_m=2.78$~\cite{footnote} that is bound to the tip of the nanorod.  Because the field strength decays very rapidly near the tips (with a characteristic length on the order of nanometers), the field is inhomogeneous over the BSA molecule.  For this reason rather than using a dipole approximation as in Eq.~(\ref{eq:freq_shift}), the frequency shift is found by numerically integrating the field intensity associated with polarizing the BSA molecule~\cite{shopova}.  The enhancement in $\delta \omega$ is the ratio of the polarization energies with and without the nanorod
\begin{equation}
\xi(R) = \frac{\int E_{nr}^*(\vec{r}) \cdot E_m(\vec{r}) \hspace{1mm} d\vec{r}}{\int E_0^*(\vec{r}) \cdot E_{0,m}(\vec{r}) \hspace{1mm} d\vec{r}}
\label{eq:enhancement}
\end{equation}
where $E_{nr}(\vec{r})$ is the field emerging from the nanorod, $E_m(\vec{r})$ is the induced field in the BSA molecule and $E_0(\vec{r})$ is a plane wave which represents the WGM field prior to the enhancement.  We consider the frequency shift due to a first-order pertubation, and therefore take the integration over the volume of the BSA molecule only~\cite{first-order}.  Fig.~\ref{fig:enhancement} (a) shows the calculated enhancement for a nanorod with $R=4$.  The shape is an asymmetric Lorentzian, on account of the plasmonic loss being non-uniform over the wavelength band.  The enhancement is very broad, with a peak value of roughly 870 on resonance ($\lambda =$ 803 nm), and decreases linearly with decreasing $R$ (Fig.~\ref{fig:enhancement} (b)).  In the inset of Fig.~\ref{fig:enhancement} (a), we show the electric field intensity inside the BSA molecule.  As a result of the inhomogeneity in $E_{nr}(\vec{r})$ near the nanorod tip, the field penetrates into the molecule very weakly, amounting to a smaller enhancement through the numerator of Eq.~(\ref{eq:enhancement}).  However, BSA is a relatively large protein~\cite{Vollmer2}, and in principle much larger enhancements could be achieved for smaller proteins.

We can now estimate the frequency shift $\delta \omega_{\rm{BSA}}$ expected for a single BSA molecule.  We consider a measurement of $\delta \omega$ with some fraction due to the binding event $\delta \omega_{\rm{BSA}}$ and the rest due to noise from fluctuations in the laser frequency $\delta \omega_{\Delta \Omega}$ and thermal fluctuations within the resonator $\delta \omega_{\Delta T}$:
\begin{equation}
\delta \omega = \delta \omega_{\rm{BSA}} + \delta \omega_{\Delta \Omega} + \delta \omega_{\Delta T}
\label{eq:bsa_freq_shift}
\end{equation}
Using Eqs.~(\ref{eq:freq_shift}) and ~(\ref{eq:enhancement}), the enhanced BSA frequency shift $\delta \omega_{\rm{BSA}} = \xi(R) \times \delta \omega_m$, where $\delta \omega_m$ is the frequency shift due to BSA prior to the enhancement.  For a standard microtoroid resonator with $V=760$ $\mu$m$^3$, $\delta \omega_m$ is about 5 kHz.  For the remaining terms in Eq.~(\ref{eq:bsa_freq_shift}), we take $\delta \omega_{\Delta \Omega}$ to be fluctuations on the order of a tunable diode laser's linewidth ($\delta \omega_{\Delta \Omega}=$ 100 kHz~\cite{Turner}) and $\delta \omega_{\Delta T}$ to be thermorefractive fluctuations in a standard microtoroid resonator.  The thermal frequency shift is given by $\delta \omega_{\Delta T}/\omega=n^{-1} \frac{dn}{dT} \Delta T$, where for SiO$_2$ $n=1.45$ is the refractive index, $\frac{dn}{dT}=1.45 \times$ 10$^{-5}$ K$^{-1}$ is the thermorefractive coefficient and $\Delta T$ is the temperature fluctuation over an averaging time $\tau$.  By following the analytical work of Gorodetsky and Grudinin~\cite{Gorodetsky} and numerically calculating the power spectral density of thermorefractive noise in a microtoroid, we found that $\Delta T = 0.6 \pm 0.1$ $\mathrm{\mu}$K for $\tau=1$ ms (see supplemental material~\cite{supp}).  However, since temperature fluctuations scale with the effective mode volume, $\Delta T \propto V_{\rm{eff}}^{-1} = \int |E(\vec{r})|^4 \hspace{1mm} d\vec{r}$, we also calculated the spectrum including the field from the nanorod $E_{nr}(\vec{r})$.  For the enhancements reported in this letter, the nanorod's contribution to thermorefractive noise was negligible.  
\begin{figure}[ht!]
	\begin{center}
	\includegraphics*[scale=0.4]{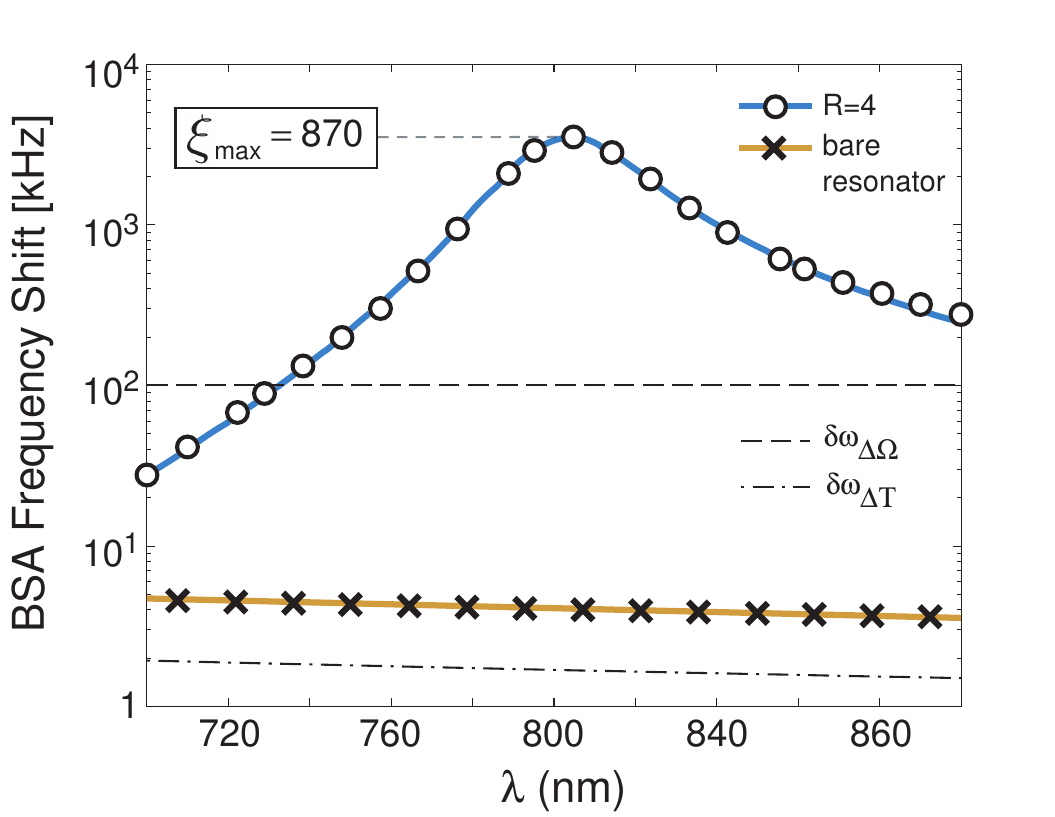}
	\caption{Predicted BSA frequency shift for a bare resonator (crosses), nanorod-enhanced resonator (circles), and typical noise contributions from frequency noise (dotted) and thermal noise (dash dotted). $V=760$ $\mu$m$^3$}
	\label{fig:final_freq_shift}%
	\end{center}
\end{figure} 

In Fig.~\ref{fig:final_freq_shift} we show the predicted BSA frequency shift for a microtoroid with $V=760$ $\mu$m$^3$.  The circles with a Lorentzian fit are the enhanced frequency shift due to a nanorod with $R=4$, and the crosses denote the shift for a bare microtoroid resonator without the plasmonic enhancement ($\xi=1$).  On resonance the enhanced frequency shift is predicted to be about 4 MHz, well above the noise contributions.  By contrast, the bare frequency shift $\delta \omega_m$ is well below the detection limit set by fluctuations in laser frequency (dashed), in agreement with another report~\cite{Lu}.  Furthermore, the thermal noise (dash dotted) is on the order of $\delta \omega_m$, suggesting that without the enhancement themorefractive noise could preclude single molecule detection for measurements with $\tau \leq$ 1 ms, even when interferometric methods are employed to cancel $\delta \omega_{\Delta \Omega}$~\cite{Knittel, Lu}.  

In summary, we have performed theoretical calulations of the LSPR excited in Au nanorods bound to a microtoroidal resonator.  The frequency shift enhancement is likely to put single molecule detection within reach.  We calculate a maximum resonant enhancement of $\xi=870$ for BSA binding to a nanorod with $R=4$, which corresponds to a frequency shift on the order of 4 MHz in a microtoroid.  

This research was funded by the Australian Research Council Grant No. DP0987146.


\begin{thebibliography}{20}%
\makeatletter
\providecommand \@ifxundefined [1]{%
 \@ifx{#1\undefined}\
}%
\providecommand \@ifnum [1]{%
 \ifnum #1\expandafter \@firstoftwo
 \else \expandafter \@secondoftwo
 \fi
}%
\providecommand \@ifx [1]{%
 \ifx #1\expandafter \@firstoftwo
 \else \expandafter \@secondoftwo
 \fi
}%
\providecommand \natexlab [1]{#1}%
\providecommand \enquote  [1]{``#1''}%
\providecommand \bibnamefont  [1]{#1}%
\providecommand \bibfnamefont [1]{#1}%
\providecommand \citenamefont [1]{#1}%
\providecommand \href@noop [0]{\@secondoftwo}%
\providecommand \href [0]{\begingroup \@sanitize@url \@href}%
\providecommand \@href[1]{\@@startlink{#1}\@@href}%
\providecommand \@@href[1]{\endgroup#1\@@endlink}%
\providecommand \@sanitize@url [0]{\catcode `\\12\catcode `\$12\catcode
  `\&12\catcode `\#12\catcode `\^12\catcode `\_12\catcode `\%12\relax}%
\providecommand \@@startlink[1]{}%
\providecommand \@@endlink[0]{}%
\providecommand \url  [0]{\begingroup\@sanitize@url \@url }%
\providecommand \@url [1]{\endgroup\@href {#1}{\urlprefix }}%
\providecommand \urlprefix  [0]{URL }%
\providecommand \Eprint [0]{\href }%
\providecommand \doibase [0]{http://dx.doi.org/}%
\providecommand \selectlanguage [0]{\@gobble}%
\providecommand \bibinfo  [0]{\@secondoftwo}%
\providecommand \bibfield  [0]{\@secondoftwo}%
\providecommand \translation [1]{[#1]}%
\providecommand \BibitemOpen [0]{}%
\providecommand \bibitemStop [0]{}%
\providecommand \bibitemNoStop [0]{.\EOS\space}%
\providecommand \EOS [0]{\spacefactor3000\relax}%
\providecommand \BibitemShut  [1]{\csname bibitem#1\endcsname}%
\let\auto@bib@innerbib\@empty

\bibitem [{\citenamefont {Cai}, \citenamefont {Painter},\ and\ \citenamefont
  {Vahala}(2000)}]{Cai}%
  \BibitemOpen
  \bibfield  {author} {\bibinfo {author} {\bibfnamefont {M.}~\bibnamefont
  {Cai}}, \bibinfo {author} {\bibfnamefont {O.}~\bibnamefont {Painter}}, \ and\
  \bibinfo {author} {\bibfnamefont {K.~J.}\ \bibnamefont {Vahala}},\ }\href
  {\doibase 10.1103/PhysRevLett.85.74} {\bibfield  {journal} {\bibinfo
  {journal} {Phys. Rev. Lett.}\ }\textbf {\bibinfo {volume} {85}},\ \bibinfo
  {pages} {74} (\bibinfo {year} {2000})}
\bibitem [{\citenamefont {D.~Armani}\ and\ \citenamefont
  {Vahala}(2003)}]{Vahala}%
  \BibitemOpen
  \bibfield  {author} {\bibinfo {author} {\bibfnamefont {S.~S.}\ \bibnamefont
  {D.~Armani}, \bibfnamefont {T.~Kippenberg}}\ and\ \bibinfo {author}
  {\bibfnamefont {K.}~\bibnamefont {Vahala}},\ }\href@noop {} {\bibfield
  {journal} {\bibinfo  {journal} {Nature}\ }\textbf {\bibinfo {volume} {421}},\
  \bibinfo {pages} {925} (\bibinfo {year} {2003})}
\bibitem [{\citenamefont {Armani}\ and\ \citenamefont
  {Vahala}(2006)}]{Heavywater}%
  \BibitemOpen
  \bibfield  {author} {\bibinfo {author} {\bibfnamefont {A.~M.}\ \bibnamefont
  {Armani}}\ and\ \bibinfo {author} {\bibfnamefont {K.~J.}\ \bibnamefont
  {Vahala}},\ }\href {\doibase 10.1364/OL.31.001896} {\bibfield  {journal}
  {\bibinfo  {journal} {Opt. Lett.}\ }\textbf {\bibinfo {volume} {31}},\
  \bibinfo {pages} {1896} (\bibinfo {year} {2006})}
\bibitem [{\citenamefont {Arnold}\ \emph {et~al.}(2003)\citenamefont {Arnold},
  \citenamefont {Khoshsima}, \citenamefont {Teraoka}, \citenamefont {Holler},\
  and\ \citenamefont {Vollmer}}]{Arnold}%
  \BibitemOpen
  \bibfield  {author} {\bibinfo {author} {\bibfnamefont {S.}~\bibnamefont
  {Arnold}}, \bibinfo {author} {\bibfnamefont {M.}~\bibnamefont {Khoshsima}},
  \bibinfo {author} {\bibfnamefont {I.}~\bibnamefont {Teraoka}}, \bibinfo
  {author} {\bibfnamefont {S.}~\bibnamefont {Holler}}, \ and\ \bibinfo {author}
  {\bibfnamefont {F.}~\bibnamefont {Vollmer}},\ }\href {\doibase
  10.1364/OL.28.000272} {\bibfield  {journal} {\bibinfo  {journal} {Opt.
  Lett.}\ }\textbf {\bibinfo {volume} {28}},\ \bibinfo {pages} {272} (\bibinfo
  {year} {2003})}
\bibitem [{\citenamefont {Knittel}\ \emph {et~al.}(2010)\citenamefont
  {Knittel}, \citenamefont {McRae}, \citenamefont {Lee},\ and\ \citenamefont
  {Bowen}}]{Knittel}%
  \BibitemOpen
  \bibfield  {author} {\bibinfo {author} {\bibfnamefont {J.}~\bibnamefont
  {Knittel}}, \bibinfo {author} {\bibfnamefont {T.~G.}\ \bibnamefont {McRae}},
  \bibinfo {author} {\bibfnamefont {K.~H.}\ \bibnamefont {Lee}}, \ and\
  \bibinfo {author} {\bibfnamefont {W.~P.}\ \bibnamefont {Bowen}},\ }\href
  {\doibase 10.1063/1.3494530} {\bibfield  {journal} {\bibinfo  {journal}
  {Applied Physics Letters}\ }\textbf {\bibinfo {volume} {97}},\ \bibinfo {eid}
  {123704} (\bibinfo {year} {2010})}
\bibitem [{\citenamefont {Armani}\ \emph {et~al.}(2007)\citenamefont {Armani},
  \citenamefont {Kulkarni}, \citenamefont {Fraser}, \citenamefont {Flagan},\
  and\ \citenamefont {Vahala}}]{Armani}%
  \BibitemOpen
  \bibfield  {author} {\bibinfo {author} {\bibfnamefont {A.~M.}\ \bibnamefont
  {Armani}}, \bibinfo {author} {\bibfnamefont {R.~P.}\ \bibnamefont
  {Kulkarni}}, \bibinfo {author} {\bibfnamefont {S.~E.}\ \bibnamefont
  {Fraser}}, \bibinfo {author} {\bibfnamefont {R.~C.}\ \bibnamefont {Flagan}},
  \ and\ \bibinfo {author} {\bibfnamefont {K.~J.}\ \bibnamefont {Vahala}},\
  }\href {\doibase 10.1126/science.1145002} {\bibfield  {journal} {\bibinfo
  {journal} {Science}\ }\textbf {\bibinfo {volume} {317}},\ \bibinfo {pages}
  {783} (\bibinfo {year} {2007})},\ \Eprint
  {} {} 
\bibitem [{\citenamefont {Vollmer}(2005)}]{Vollmer2}%
  \BibitemOpen
  \bibfield  {author} {\bibinfo {author} {\bibfnamefont {F.}~\bibnamefont
  {Vollmer}},\ }\href@noop {} {\bibfield  {journal} {\bibinfo  {journal}
  {B.I.F. Futura}\ }\textbf {\bibinfo {volume} {20}},\ \bibinfo {pages} {239}
  (\bibinfo {year} {2005})}
\bibitem [{\citenamefont {Lu}\ \emph {et~al.}(2010)\citenamefont {Lu},
  \citenamefont {Lee}, \citenamefont {Cheni}, \citenamefont {Herchaki},
  \citenamefont {Kim},\ and\ \citenamefont {Vahala}}]{Lu}%
  \BibitemOpen
  \bibfield  {author} {\bibinfo {author} {\bibfnamefont {T.}~\bibnamefont
  {Lu}}, \bibinfo {author} {\bibfnamefont {H.}~\bibnamefont {Lee}}, \bibinfo
  {author} {\bibfnamefont {T.}~\bibnamefont {Cheni}}, \bibinfo {author}
  {\bibfnamefont {S.}~\bibnamefont {Herchaki}}, \bibinfo {author}
  {\bibfnamefont {J.}~\bibnamefont {Kim}}, \ and\ \bibinfo {author}
  {\bibfnamefont {K.}~\bibnamefont {Vahala}},\ }in\ \href@noop {} {\emph
  {\bibinfo {booktitle} {Frontiers in Optics, OSA Technical Digest}}}\
  (\bibinfo {year} {2010})\ p.\ \bibinfo {pages} {PDPB5}
\bibitem [{\citenamefont {Shopova}\ \emph {et~al.}(2011)\citenamefont
  {Shopova}, \citenamefont {Rajmangal}, \citenamefont {Holler},\ and\
  \citenamefont {Arnold}}]{shopova}%
  \BibitemOpen
  \bibfield  {author} {\bibinfo {author} {\bibfnamefont {S.~I.}\ \bibnamefont
  {Shopova}}, \bibinfo {author} {\bibfnamefont {R.}~\bibnamefont {Rajmangal}},
  \bibinfo {author} {\bibfnamefont {S.}~\bibnamefont {Holler}}, \ and\ \bibinfo
  {author} {\bibfnamefont {S.}~\bibnamefont {Arnold}},\ }\href {\doibase
  10.1063/1.3599584} {\bibfield  {journal} {\bibinfo  {journal} {Applied
  Physics Letters}\ }\textbf {\bibinfo {volume} {98}},\ \bibinfo {eid} {243104}
  (\bibinfo {year} {2011})}
\bibitem [{\citenamefont {Santiago-Cordoba}\ \emph {et~al.}(2011)\citenamefont
  {Santiago-Cordoba}, \citenamefont {Boriskina}, \citenamefont {Vollmer},\ and\
  \citenamefont {Demirel}}]{Vollmer-NP}%
  \BibitemOpen
  \bibfield  {author} {\bibinfo {author} {\bibfnamefont {M.~A.}\ \bibnamefont
  {Santiago-Cordoba}}, \bibinfo {author} {\bibfnamefont {S.~V.}\ \bibnamefont
  {Boriskina}}, \bibinfo {author} {\bibfnamefont {F.}~\bibnamefont {Vollmer}},
  \ and\ \bibinfo {author} {\bibfnamefont {M.~C.}\ \bibnamefont {Demirel}},\
  }\href {\doibase 10.1063/1.3599706} {\bibfield  {journal} {\bibinfo
  {journal} {Applied Physics Letters}\ }\textbf {\bibinfo {volume} {99}},\
  \bibinfo {eid} {073701} (\bibinfo {year} {2011})}
\bibitem [{\citenamefont {Grzelczak}\ \emph {et~al.}(2007)\citenamefont
  {Grzelczak}, \citenamefont {Pérez-Juste}, \citenamefont {García~de Abajo},\
  and\ \citenamefont {Liz-Marzán}}]{pt-nanorods}%
  \BibitemOpen
  \bibfield  {author} {\bibinfo {author} {\bibfnamefont {M.}~\bibnamefont
  {Grzelczak}}, \bibinfo {author} {\bibfnamefont {J.}~\bibnamefont 
  {Pérez-Juste}}, \bibinfo {author} {\bibfnamefont {F.~J.}\ \bibnamefont
  {García~de Abajo}}, \ and\ \bibinfo {author} {\bibfnamefont {L.~M.}\
  \bibnamefont {Liz-Marzán}},\ }\href {\doibase 10.1021/jp0671502} {\bibfield
  {journal} {\bibinfo  {journal} {The Journal of Physical Chemistry C}\
  }\textbf {\bibinfo {volume} {111}},\ \bibinfo {pages} {6183} (\bibinfo {year}
  {2007})},\ \Eprint
  {}
  {} 
\bibitem [{\citenamefont {Garcia de Abajo}\ and\ \citenamefont
  {Howie}(1998)}]{Abajo1}%
  \BibitemOpen
  \bibfield  {author} {\bibinfo {author} {\bibfnamefont {F.~J.}\ \bibnamefont
  {Garcia de Abajo}}\ and\ \bibinfo {author} {\bibfnamefont
  {A.}~\bibnamefont {Howie}},\ }\href {\doibase 10.1103/PhysRevLett.80.5180}
  {\bibfield  {journal} {\bibinfo  {journal} {Phys. Rev. Lett.}\ }\textbf
  {\bibinfo {volume} {80}},\ \bibinfo {pages} {5180} (\bibinfo {year}
  {1998})}
\bibitem [{\citenamefont {Garcia de Abajo}\ and\ \citenamefont 
  {Howie}(2002)}]{Abajo2}%
  \BibitemOpen
  \bibfield  {author} {\bibinfo {author} {\bibfnamefont {F.~J.}\ \bibnamefont
  {Garcia de Abajo}}\ and\ \bibinfo {author} {\bibfnamefont
  {A.}~\bibnamefont {Howie}},\ }\href {\doibase 10.1103/PhysRevB.65.115418}
  {\bibfield  {journal} {\bibinfo  {journal} {Phys. Rev. B}\ }\textbf {\bibinfo
  {volume} {65}},\ \bibinfo {pages} {115418} (\bibinfo {year}
  {2002})}
\bibitem [{\citenamefont {Johnson}\ and\ \citenamefont
  {Christy}(1972)}]{Johnson}%
  \BibitemOpen
  \bibfield  {author} {\bibinfo {author} {\bibfnamefont {P.~B.}\ \bibnamefont
  {Johnson}}\ and\ \bibinfo {author} {\bibfnamefont {R.~W.}\ \bibnamefont
  {Christy}},\ }\href {\doibase 10.1103/PhysRevB.6.4370} {\bibfield  {journal}
  {\bibinfo  {journal} {Phys. Rev. B}\ }\textbf {\bibinfo {volume} {6}},\
  \bibinfo {pages} {4370} (\bibinfo {year} {1972})}
\bibitem [{\citenamefont {Kreibig}\ and\ \citenamefont
  {Vollmer}(1995)}]{Kreibig}%
  \BibitemOpen
  \bibfield  {author} {\bibinfo {author} {\bibfnamefont {U.}~\bibnamefont
  {Kreibig}}\ and\ \bibinfo {author} {\bibfnamefont {M.}~\bibnamefont
  {Vollmer}},\ }\href@noop {} {\emph {\bibinfo {title} {Optical properties of
  metal clusters}}}\ (\bibinfo  {publisher} {Springer},\ \bibinfo {address}
  {Berlin, Germany},\ \bibinfo {year} {1995})
\bibitem [{\citenamefont {Mazzei}\ \emph {et~al.}(2007)\citenamefont {Mazzei},
  \citenamefont {G\"otzinger}, \citenamefont {de~S.~Menezes}, \citenamefont
  {Zumofen}, \citenamefont {Benson},\ and\ \citenamefont
  {Sandoghdar}}]{Mazzei}%
  \BibitemOpen
  \bibfield  {author} {\bibinfo {author} {\bibfnamefont {A.}~\bibnamefont
  {Mazzei}}, \bibinfo {author} {\bibfnamefont {S.}~\bibnamefont {G\"otzinger}},
  \bibinfo {author} {\bibfnamefont {L.}~\bibnamefont {de~S.~Menezes}}, \bibinfo
  {author} {\bibfnamefont {G.}~\bibnamefont {Zumofen}}, \bibinfo {author}
  {\bibfnamefont {O.}~\bibnamefont {Benson}}, \ and\ \bibinfo {author}
  {\bibfnamefont {V.}~\bibnamefont {Sandoghdar}},\ }\href {\doibase
  10.1103/PhysRevLett.99.173603} {\bibfield  {journal} {\bibinfo  {journal}
  {Phys. Rev. Lett.}\ }\textbf {\bibinfo {volume} {99}},\ \bibinfo {pages}
  {173603} (\bibinfo {year} {2007})}
\bibitem [{\citenamefont {Zhu}\ \emph {et~al.}(2010)\citenamefont {Zhu},
  \citenamefont {Ozdemir}, \citenamefont {Xiao}, \citenamefont {Lin},
  \citenamefont {He}, \citenamefont {Chen},\ and\ \citenamefont {Yang}}]{Yang}%
  \BibitemOpen
  \bibfield  {author} {\bibinfo {author} {\bibfnamefont {J.}~\bibnamefont
  {Zhu}}, \bibinfo {author} {\bibfnamefont {S.~K.}\ \bibnamefont {Ozdemir}},
  \bibinfo {author} {\bibfnamefont {Y.}~\bibnamefont {Xiao}}, \bibinfo {author}
  {\bibfnamefont {L.}~\bibnamefont {Lin}}, \bibinfo {author} {\bibfnamefont
  {L.}~\bibnamefont {He}}, \bibinfo {author} {\bibfnamefont {D.}~\bibnamefont
  {Chen}}, \ and\ \bibinfo {author} {\bibfnamefont {L.}~\bibnamefont {Yang}},\
  }\href
  {http://www.nature.com/nphoton/journal/v4/n1/suppinfo/nphoton.2009.237_S1.ht%
ml} {\bibfield  {journal} {\bibinfo  {journal} {Nature Photonics}\ }\textbf
  {\bibinfo {volume} {4}} (\bibinfo {year} {2010})}
\bibitem [{\citenamefont {Vollmer}(2005)}]{footnote}%
  \BibitemOpen
  \bibfield  {author} {\bibinfo {author} }\href@noop {} {\bibfield  {journal} {\bibinfo  {journal}
  {Using the polarizability for BSA from Ref. [7] and a radius of 3 nm, we use the standard formula for polarizability (see J. Jackson, \emph{Classical Electrodynamics} (Wiley, New York, 1999)) to obtain a relative permittivity of 2.78 for BSA}\ } }
\bibitem [{\citenamefont {Teraoka}\ and\ \citenamefont
  {Arnold}(2006)}]{first-order}%
  \BibitemOpen
  \bibfield  {author} {\bibinfo {author} {\bibfnamefont {I.}~\bibnamefont
  {Teraoka}}\ and\ \bibinfo {author} {\bibfnamefont {S.}~\bibnamefont
  {Arnold}},\ }\href {\doibase 10.1364/JOSAB.23.001381} {\bibfield  {journal}
  {\bibinfo  {journal} {J. Opt. Soc. Am. B}\ }\textbf {\bibinfo {volume}
  {23}},\ \bibinfo {pages} {1381} (\bibinfo {year} {2006})}
\bibitem [{\citenamefont {Turner}\ \emph {et~al.}(2002)\citenamefont {Turner},
  \citenamefont {Weber}, \citenamefont {Hawthorn},\ and\ \citenamefont
  {Scholten}}]{Turner}%
  \BibitemOpen
  \bibfield  {author} {\bibinfo {author} {\bibfnamefont {L.~D.}\ \bibnamefont
  {Turner}}, \bibinfo {author} {\bibfnamefont {K.~P.}\ \bibnamefont {Weber}},
  \bibinfo {author} {\bibfnamefont {C.~J.}\ \bibnamefont {Hawthorn}}, \ and\
  \bibinfo {author} {\bibfnamefont {R.~E.}\ \bibnamefont {Scholten}},\ }\href
  {\doibase DOI: 10.1016/S0030-4018(01)01689-3} {\bibfield  {journal} {\bibinfo
   {journal} {Optics Communications}\ }\textbf {\bibinfo {volume} {201}},\
  \bibinfo {pages} {391 } (\bibinfo {year} {2002})}
\bibitem [{\citenamefont {Gorodetsky}\ and\ \citenamefont
  {Grudinin}(2004)}]{Gorodetsky}%
  \BibitemOpen
  \bibfield  {author} {\bibinfo {author} {\bibfnamefont {M.~L.}\ \bibnamefont
  {Gorodetsky}}\ and\ \bibinfo {author} {\bibfnamefont {I.~S.}\ \bibnamefont
  {Grudinin}},\ }\href {\doibase 10.1364/JOSAB.21.000697} {\bibfield  {journal}
  {\bibinfo  {journal} {J. Opt. Soc. Am. B}\ }\textbf {\bibinfo {volume}
  {21}},\ \bibinfo {pages} {697} (\bibinfo {year} {2004})}
\bibitem [{\citenamefont {See supplementary material at [URL will be inserted
  by AIP]}()}]{supp}%
  \BibitemOpen
  \bibfield  {author} {\bibinfo {author} {\bibfnamefont {} \bibnamefont {See supplementary material at [URL will be
  inserted by AIP]}}\ }\href@noop {} {}







\end{thebibliography}

%

\end{document}